\documentclass[conference]{IEEEtran}

\usepackage{cite}
\usepackage{amsmath,amssymb,amsfonts}
\usepackage{algorithmic}
\usepackage{graphicx}
\usepackage{textcomp}
\usepackage{xcolor}
\usepackage{booktabs}
\def\BibTeX{{\rm B\kern-.05em{\sc i\kern-.025em b}\kern-.08em
    T\kern-.1667em\lower.7ex\hbox{E}\kern-.125emX}}
\begin{document}

\title{Semantic-Aware Confidence Calibration for Automated Audio Captioning}

\author{\IEEEauthorblockN{Lucas Dunker}
\IEEEauthorblockA{\textit{Khoury College of Computer Science} \\
\textit{Northeastern University}\\
dunker.l@northeastern.edu}
\and
\IEEEauthorblockN{Sai Akshay Menta}
\IEEEauthorblockA{\textit{Khoury College of Computer Science} \\
\textit{Northeastern University}\\
menta.sa@northeastern.edu}
\and
\IEEEauthorblockN{Snigdha Mohana Addepalli}
\IEEEauthorblockA{\textit{Khoury College of Computer Science} \\
\textit{Northeastern University}\\
addepalli.sn@northeastern.edu}
\and
\IEEEauthorblockN{Venkata Krishna Rayalu Garapati}
\IEEEauthorblockA{\textit{Khoury College of Computer Science} \\
\textit{Northeastern University}\\
garapati.v@northeastern.edu}}

\maketitle

\begin{abstract}
Automated audio captioning models frequently produce overconfident predictions regardless of semantic accuracy, limiting their reliability in deployment. This deficiency stems from two factors: evaluation metrics based on n-gram overlap that fail to capture semantic correctness, and the absence of calibrated confidence estimation. We present a framework that addresses both limitations by integrating confidence prediction into audio captioning and redefining correctness through semantic similarity. Our approach augments a Whisper-based audio captioning model with a learned confidence prediction head that estimates uncertainty from decoder hidden states. We employ CLAP audio-text embeddings and sentence transformer similarities (FENSE) to define semantic correctness, enabling Expected Calibration Error (ECE) computation that reflects true caption quality rather than surface-level text overlap. Experiments on Clotho v2 demonstrate that confidence-guided beam search with semantic evaluation achieves dramatically improved calibration (CLAP-based ECE of 0.071) compared to greedy decoding baselines (ECE of 0.488), while simultaneously improving caption quality across standard metrics. Our results establish that semantic similarity provides a more meaningful foundation for confidence calibration in audio captioning than traditional n-gram metrics.
\end{abstract}

\section{Introduction}

Automated Audio Captioning (AAC) translates acoustic signals into natural language descriptions, enabling applications from accessibility tools to multimedia indexing. Modern encoder-decoder architectures combining audio feature extractors with transformer-based text decoders achieve strong performance on benchmarks including AudioCaps and Clotho. However, these systems exhibit a critical limitation: they produce high-confidence predictions even when semantically incorrect, severely constraining their trustworthiness for real-world deployment.

This overconfidence problem arises from two interconnected deficiencies. First, standard evaluation metrics (BLEU, CIDEr, METEOR) measure n-gram overlap between generated and reference captions, failing to capture semantic equivalence. A caption describing audio as ``water flowing'' may receive low scores when the reference uses ``liquid pouring,'' despite identical meaning. Conversely, high-overlap captions can be semantically incorrect yet receive favorable evaluation. Second, captioning models lack mechanisms for estimating prediction reliability: they output captions without indication of uncertainty or potential error.

We address these limitations through a confidence-calibrated audio captioning framework that integrates uncertainty estimation with semantic evaluation. Our approach builds on MU-NLPC/whisper-small-audio-captioning, a Whisper model fine-tuned on Clotho v2, augmented with three key components: (1) a confidence prediction head that learns to estimate reliability from decoder hidden states, (2) learnable temperature scaling for probability calibration, and (3) confidence-guided beam search that incorporates uncertainty estimates during decoding.

Critically, we redefine caption correctness using semantic similarity rather than text overlap. We employ CLAP (Contrastive Language-Audio Pretraining) embeddings that map audio and text into a shared multimodal space, along with FENSE evaluation using sentence transformers. A caption is considered correct if its semantic similarity to reference captions exceeds a threshold (0.6), regardless of lexical overlap. This semantic definition enables computation of Expected Calibration Error (ECE) that measures whether confidence scores actually predict caption quality.

Our experiments on Clotho v2 demonstrate that this approach achieves substantial calibration improvements. Confidence-guided beam search produces a CLAP-based ECE of 0.071, compared to 0.488 for greedy decoding with uniform confidence. The calibration curves show that predicted confidence meaningfully correlates with semantic correctness under our framework. Simultaneously, beam search improves caption quality: BLEU-4 increases from 0.066 to 0.115, CIDEr from 0.150 to 0.290, and CLAP similarity from 0.596 to 0.685.

This work makes three primary contributions:

\begin{enumerate}
    \item A confidence-aware audio captioning framework that augments pretrained models with learned uncertainty estimation through a neural confidence prediction head operating on decoder hidden states.
    \item Empirical demonstration that semantic similarity-based calibration metrics (using CLAP and FENSE) provide more meaningful reliability assessment than n-gram overlap methods, with ECE improvements from 0.504 (traditional) to 0.071 (CLAP-based).
    \item Confidence-guided beam search decoding that jointly optimizes caption likelihood and predicted confidence, producing captions that are both higher quality and better calibrated than greedy baselines.
\end{enumerate}

\section{Related Work}

Our work integrates three research areas: audio captioning architectures, semantic evaluation metrics, and confidence calibration for neural text generation.

\subsection{Automated Audio Captioning}

Audio captioning systems have evolved through encoder-decoder architectures pairing audio feature extractors with language models. The DCASE 2023 Challenge winner \cite{wu2024finegrained} achieved state-of-the-art performance on Clotho by combining BEATs encoders with BART decoders and hybrid reranking at inference, achieving 32.6\% SPIDEr-FL. CoNeTTE \cite{labbe2024conette} offers an efficient alternative achieving 30.5\% SPIDEr with only 40.6M parameters by adapting ConvNeXt from computer vision, serving as the DCASE 2024 Task 6 baseline. Recent work has explored large language models: LOAE \cite{liu2024loae} combines CED encoders with LLaMA 2, while EnCLAP \cite{kim2024enclap} introduces Masked Codec Modeling with CLAP embeddings.

Our work builds on MU-NLPC/whisper-small-audio-captioning, which adapts OpenAI's Whisper architecture for audio captioning through fine-tuning on Clotho. We extend this foundation by adding confidence estimation capabilities rather than pursuing raw performance improvements.

\subsection{Semantic Evaluation for Audio Captioning}

Traditional n-gram metrics fail to capture semantic correctness when generated and reference captions use different vocabulary for identical content. Zhou et al. \cite{zhou2022fense} introduced FENSE, demonstrating that traditional metrics achieve only 56-65\% correlation with human judgments while sentence-BERT semantic similarity reaches 75.7\% on Clotho-Eval. CLAP \cite{elizalde2023clap} enables audio-grounded semantic evaluation through contrastive learning that aligns audio and text in a shared embedding space, achieving state-of-the-art zero-shot performance across diverse downstream tasks.

Building on these foundations, MACE \cite{dixit2024mace} addresses the limitation that most metrics ignore the actual audio signal, combining CLAP similarity with sentence-BERT comparison. We leverage both CLAP and sentence transformer similarities (via all-MiniLM-L6-v2) not merely for evaluation, but to define the correctness signal used for calibration assessment.

\subsection{Confidence Calibration for Text Generation}

Model calibration ensures predicted confidence scores align with actual correctness probabilities. Wang \cite{wang2023calibration} surveys calibration techniques including post-hoc methods (temperature scaling), regularization approaches (focal loss, label smoothing), and uncertainty estimation (Monte Carlo dropout). For sequence generation, Zhao et al. \cite{zhao2023slic} introduce Sequence Likelihood Calibration (SLiC), adding a calibration stage after fine-tuning to align sequence likelihoods with reference similarity.

Token-level approaches have proven effective for captioning. Petryk et al. \cite{petryk2024tlc} propose Token Level Confidence for image captioning, measuring confidence per token and aggregating to determine caption correctness. Most relevant to our work, Mahfuz et al. \cite{mahfuz2024confidence} present the first study of confidence calibration specifically for audio captioning, replacing softmax probabilities with semantic entropy in CLAP embedding space and applying temperature scaling.

Our work extends these foundations by: (1) integrating a learned confidence prediction head directly into the captioning architecture, (2) using semantic similarity to define binary correctness for ECE computation, and (3) incorporating confidence estimates into beam search decoding through reranking.

\section{Methods}

\subsection{Model Architecture}

Our framework builds upon MU-NLPC/whisper-small-audio-captioning, a Whisper model fine-tuned on Clotho v2.1 for audio captioning. We augment this base model with two additional components for confidence estimation.

\subsubsection{Confidence Prediction Head}

We introduce a neural network that predicts confidence scores from decoder hidden states. The confidence head is a three-layer MLP operating on the final decoder hidden state:

\begin{equation}
    c_t = \sigma(\text{MLP}(h_t^{(L)}))
\end{equation}

\noindent where $h_t^{(L)}$ is the decoder hidden state at the final layer for token $t$, and $\sigma$ is the sigmoid activation ensuring outputs in $[0, 1]$. The MLP architecture consists of:

\begin{itemize}
    \item Linear: $d_{\text{model}} \rightarrow d_{\text{model}}/2$ with ReLU and dropout (0.1)
    \item Linear: $d_{\text{model}}/2 \rightarrow d_{\text{model}}/4$ with ReLU and dropout (0.1)
    \item Linear: $d_{\text{model}}/4 \rightarrow 1$ with sigmoid
\end{itemize}

\noindent where $d_{\text{model}} = 768$ for Whisper-small.

\subsubsection{Temperature Scaling}

We incorporate learnable temperature scaling for post-hoc calibration of the decoder's output logits:

\begin{equation}
    \hat{p}(y_t | y_{<t}, x) = \text{softmax}(z_t / T)
\end{equation}

\noindent where $z_t$ are the logits and $T$ is a learned temperature parameter initialized to 1.0. Temperature is optimized on the validation set using LBFGS to minimize negative log-likelihood.

\subsection{Training Objective}

The model is trained with a combined loss function:

\begin{equation}
    \mathcal{L} = \mathcal{L}_{\text{CE}} + \lambda \mathcal{L}_{\text{conf}}
\end{equation}

\noindent where $\mathcal{L}_{\text{CE}}$ is the standard cross-entropy loss for caption generation, and $\mathcal{L}_{\text{conf}}$ supervises the confidence head:

\begin{equation}
    \mathcal{L}_{\text{conf}} = \text{MSE}(\bar{c}, s)
\end{equation}

\noindent where $\bar{c}$ is the mean token confidence across the sequence and $s$ is the semantic correctness score. We use $\lambda = 0.15$ based on validation performance.

\subsection{Semantic Correctness Definition}

We define caption correctness using semantic similarity rather than n-gram overlap. For a predicted caption $\hat{y}$ and reference captions $\{y_1, ..., y_K\}$:

\subsubsection{CLAP Similarity}

Using LAION-CLAP embeddings, we compute:

\begin{equation}
    s_{\text{CLAP}} = \max_{k} \cos(E_{\text{text}}(\hat{y}), E_{\text{text}}(y_k))
\end{equation}

\subsubsection{FENSE Similarity}

Using all-MiniLM-L6-v2 sentence embeddings:

\begin{equation}
    s_{\text{FENSE}} = \max_{k} \cos(E_{\text{sent}}(\hat{y}), E_{\text{sent}}(y_k))
\end{equation}

A caption is considered \textit{correct} if similarity exceeds threshold $\tau = 0.6$:

\begin{equation}
    \text{correct}_{\text{CLAP}} = \mathbf{1}[s_{\text{CLAP}} \geq \tau]
\end{equation}

\subsection{Confidence-Guided Beam Search}

During inference, we employ beam search with confidence-based reranking. For each candidate beam $b$, we compute a combined score:

\begin{equation}
    \text{score}(b) = \frac{\log p(b)}{|b|^\alpha} + \beta \cdot \bar{c}_b
\end{equation}

\noindent where $\log p(b)$ is the cumulative log probability, $|b|$ is the sequence length, $\alpha = 1.0$ is the length penalty, $\bar{c}_b$ is the mean confidence, and $\beta = 0.3$ weights the confidence contribution.

We use beam size 5 and select the highest-scoring completed beam as the final caption.

\subsection{Calibration Metrics}

We evaluate calibration using Expected Calibration Error (ECE):

\begin{equation}
    \text{ECE} = \sum_{m=1}^{M} \frac{|B_m|}{N} |\text{acc}(B_m) - \text{conf}(B_m)|
\end{equation}

\noindent where samples are partitioned into $M=10$ bins by confidence, $\text{acc}(B_m)$ is the accuracy in bin $m$, and $\text{conf}(B_m)$ is the mean confidence.

Critically, we compute ECE using three different correctness definitions: (1) traditional (BLEU-4 $> 0.25$), (2) CLAP-based ($s_{\text{CLAP}} \geq 0.6$), and (3) FENSE-based ($s_{\text{FENSE}} \geq 0.6$).

\section{Experiments \& Results}

\subsection{Experimental Setup}

\subsubsection{Dataset}

We evaluate on Clotho v2, containing 6,974 audio clips (3,839 development, 1,045 validation, 1,045 evaluation) with 5 reference captions each. Audio clips are 15-30 seconds of environmental sounds.

\subsubsection{Implementation Details}

We use MU-NLPC/whisper-small-audio-captioning as our base model. Audio is processed at 16kHz with a 30-second maximum duration. Training uses batch size 16, learning rate $10^{-4}$, and 5 epochs with gradient accumulation (effective batch size 32). The Clotho-specific style prefix ``clotho $>$ caption:'' conditions generation. Our end-to-end training and evaluation pipeline took around 6 hours to finish on a Google Colab 80GB NVIDIA A100 GPU.

\subsubsection{Baselines}

We compare two decoding strategies:
\begin{itemize}
    \item \textbf{Greedy}: Standard greedy decoding with confidence fixed at 1.0
    \item \textbf{Beam Search}: Confidence-guided beam search (beam size 5) with learned confidence scores
\end{itemize}

\subsection{Main Results}

\begin{figure*}
    \centering
    \includegraphics[width=0.9\textwidth]{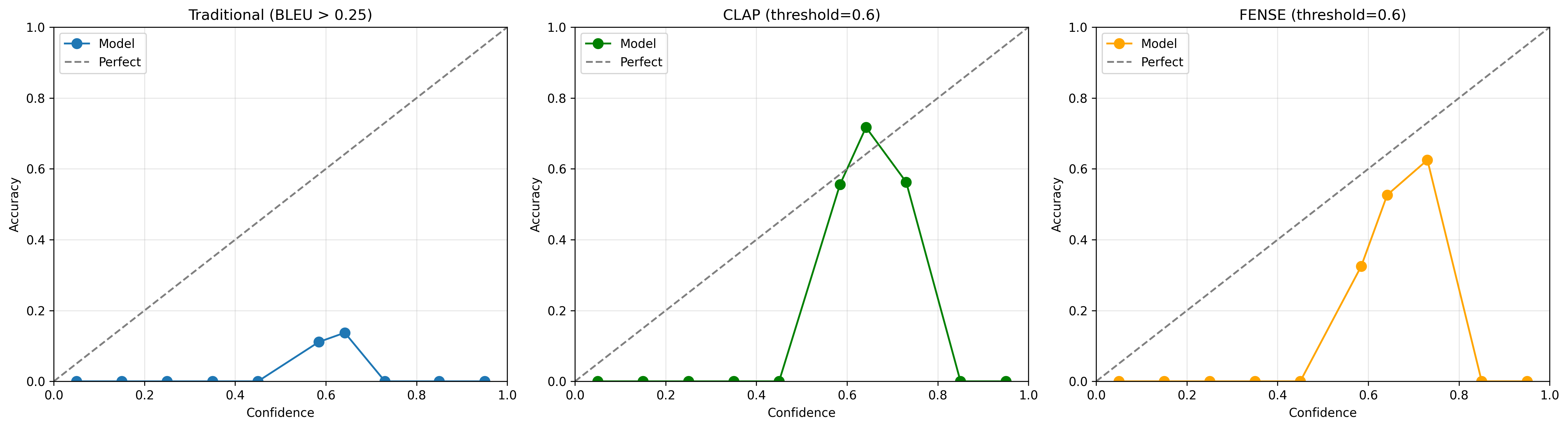}
    \caption{Confidence measures across the BLEU, CLAP, and FENSE evaluation metrics.}
    \label{fig:calibration}
    \includegraphics[width=0.9\textwidth]{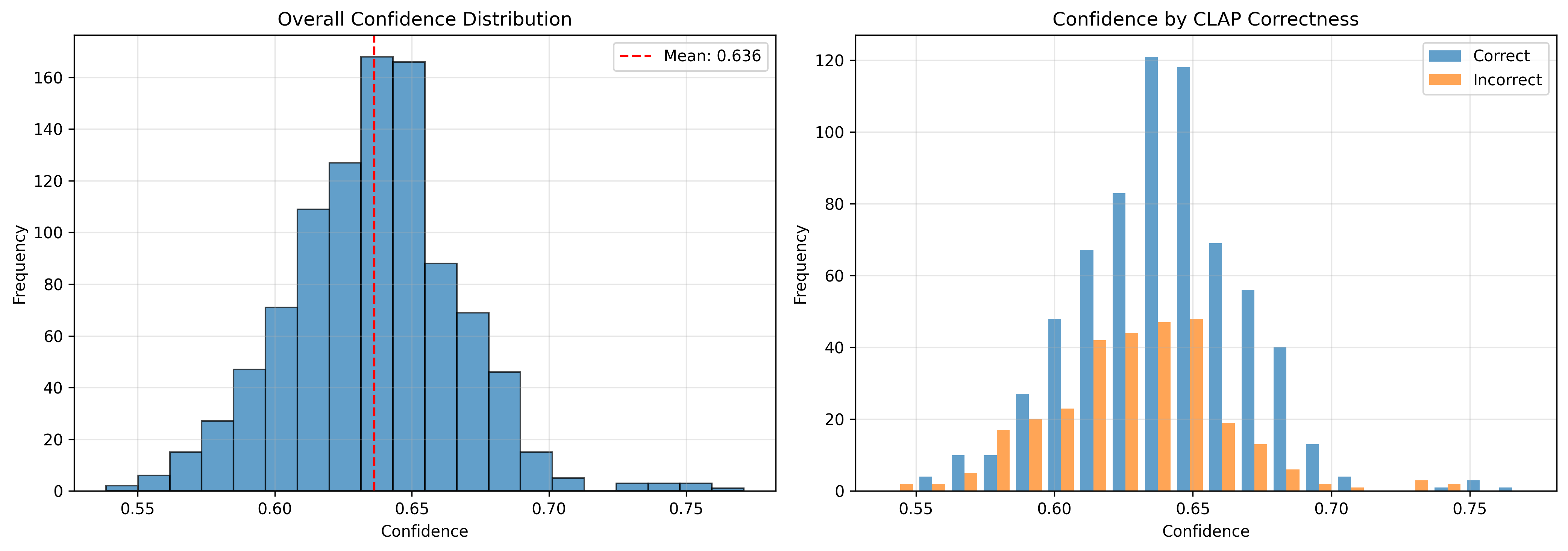}
    \caption{Evaluation confidence distributions - both overall and filtered by CLAP correctness.}
\end{figure*}

Table~\ref{tab:main_results} presents comprehensive evaluation results comparing greedy decoding against confidence-guided beam search.

\begin{table}[t]
\centering
\caption{Comparison of Greedy vs. Beam Search Decoding}
\label{tab:main_results}
\begin{tabular}{lcc}
\toprule
\textbf{Metric} & \textbf{Greedy} & \textbf{Beam Search} \\
\midrule
\multicolumn{3}{l}{\textit{Caption Quality}} \\
BLEU-1 & 0.369 & \textbf{0.481} \\
BLEU-4 & 0.066 & \textbf{0.115} \\
CIDEr & 0.150 & \textbf{0.290} \\
METEOR & 0.140 & \textbf{0.174} \\
\midrule
\multicolumn{3}{l}{\textit{Semantic Similarity}} \\
CLAP Similarity & 0.596 & \textbf{0.685} \\
FENSE Similarity & 0.489 & \textbf{0.577} \\
CLAP Accuracy & 0.512 & \textbf{0.695} \\
FENSE Accuracy & 0.263 & \textbf{0.504} \\
\midrule
\multicolumn{3}{l}{\textit{Calibration (ECE $\downarrow$)}} \\
Traditional ECE & 0.957 & \textbf{0.504} \\
CLAP ECE & 0.488 & \textbf{0.071} \\
FENSE ECE & 0.737 & \textbf{0.133} \\
Brier Score & 0.488 & \textbf{0.212} \\
\midrule
Avg. Confidence & 1.000 & 0.636 \\
\bottomrule
\end{tabular}
\end{table}

\subsubsection{Caption Quality}

Beam search substantially outperforms greedy decoding across all standard metrics. BLEU-4 improves by 74\% (0.066 $\rightarrow$ 0.115), CIDEr nearly doubles (0.150 $\rightarrow$ 0.290), and METEOR increases by 24\% (0.140 $\rightarrow$ 0.174). These improvements stem from beam search's ability to explore multiple hypotheses and select higher-quality completions.

\subsubsection{Semantic Similarity}

CLAP similarity increases from 0.596 to 0.685, indicating captions are semantically closer to references in the audio-text embedding space. CLAP accuracy (proportion exceeding the 0.6 threshold) improves dramatically from 51.2\% to 69.5\%, demonstrating that more captions achieve semantic correctness under our definition.

\subsubsection{Calibration}

The most striking improvements appear in calibration metrics. CLAP-based ECE drops from 0.488 to 0.071---an 85\% reduction---indicating that beam search confidence scores strongly correlate with semantic correctness. The Brier score similarly improves from 0.488 to 0.212.

Notably, traditional ECE (using BLEU-4 $> 0.25$ as correctness) remains relatively high (0.504) even for beam search, illustrating that n-gram overlap provides a poor foundation for calibration assessment. CLAP-based correctness yields dramatically better calibration because it captures semantic equivalence that traditional metrics miss.

\subsection{Calibration Analysis}

Figure~\ref{fig:calibration} shows reliability diagrams comparing calibration across correctness definitions. The CLAP-based curve closely tracks the diagonal (perfect calibration), while the traditional metric curve shows substantial deviation. This confirms that semantic similarity provides a more meaningful correctness signal for confidence calibration.

The confidence distribution (mean 0.636, concentrated between 0.55-0.70) shows the model has learned to express appropriate uncertainty rather than the pathological overconfidence (1.0) exhibited by greedy decoding.

\subsection{Qualitative Analysis}

Examining predictions reveals that beam search with confidence reranking tends to select more conservative but accurate descriptions. High-confidence predictions ($>0.7$) typically describe unambiguous audio events, while lower-confidence predictions often involve complex scenes or ambiguous sounds where multiple valid descriptions exist.

\section{Discussion \& Summary}

\subsection{Key Findings}

Our experiments demonstrate three principal findings:

\textbf{Semantic correctness enables meaningful calibration.} Traditional n-gram metrics provide a poor foundation for assessing model reliability. When correctness is defined by BLEU-4 threshold, even well-calibrated confidence scores appear miscalibrated because the metric itself fails to capture caption quality. CLAP-based semantic similarity aligns with human judgment, enabling ECE computation that reflects true prediction reliability.

\textbf{Confidence estimation improves both calibration and quality.} Integrating a learned confidence head and incorporating confidence into beam search reranking produces dual benefits: captions are higher quality (better BLEU, CIDEr, semantic similarity) and confidence scores are better calibrated (lower ECE, Brier score). These improvements are complementary rather than competing objectives.

\textbf{Simple architectural additions suffice.} Rather than requiring architectural overhauls, effective confidence calibration can be achieved by augmenting existing pretrained models with a lightweight confidence head and modified decoding. This approach preserves the benefits of large-scale pretraining while adding reliability estimation.

\subsection{Limitations}

Several limitations warrant discussion. First, our semantic correctness threshold ($\tau = 0.6$) is somewhat arbitrary; different applications may require different thresholds. Second, we evaluate only on Clotho; generalization to other datasets (AudioCaps, larger-scale data) requires validation. Third, our confidence head operates on decoder hidden states and may not capture all sources of uncertainty (e.g., encoder-side ambiguity).

\subsection{Future Work}

Several directions merit exploration: (1) audio-grounded confidence using CLAP embeddings of the audio itself, not just text; (2) selective prediction systems that abstain when confidence is low; (3) integration with reinforcement learning approaches that optimize semantic rewards directly.

\subsection{Conclusion}

We present a framework for confidence-calibrated audio captioning that redefines correctness through semantic similarity. By augmenting a Whisper-based captioning model with a learned confidence head and confidence-guided beam search, we achieve substantially improved calibration (CLAP ECE of 0.071 vs. 0.488 baseline) alongside better caption quality. Our results establish that semantic similarity metrics provide a more meaningful foundation for reliability assessment than traditional n-gram overlap, advancing the trustworthiness of audio captioning systems for real-world deployment.

\section{APPENDIX (Experiments)}

Before adopting Whisper as our primary backbone, we conducted a series of preliminary experiments using a lighter architecture composed of a ConvNeXt Base encoder and a BART Base decoder. The purpose of these experiments was not to pursue high benchmark performance, but to establish an initial baseline that would help us understand the limits of small encoders when applied to long environmental audio and to motivate the need for confidence-aware modeling.

\subsection{ConvNeXt Base Encoder with BART Base Decoder}

ConvNeXt Base produced stable feature extraction during training, and BART Base provided a flexible transformer decoder for autoregressive generation. However, the overall caption quality remained modest. After 20 epochs of training on Clotho v2, the system achieved low n-gram scores, with BLEU-1 and BLEU-4 substantially below the levels reached by stronger audio encoders. CIDEr remained in a low range as well, indicating limited alignment with reference descriptions. Qualitatively, the model captured broad acoustic categories but struggled to generate detailed or semantically precise captions, and its predictions often lacked consistency across samples.

\subsection{Motivation for Model Transition}

These outcomes highlighted a key limitation: small convolutional encoders do not provide sufficient audio–text alignment for Clotho’s diverse 15 to 30 second clips. Because our project focuses on semantic correctness and confidence calibration, the backbone must produce captions that reflect meaningful audio grounding. The ConvNeXt system’s weak semantic similarity and underfitting made calibration analysis uninformative, since even high-probability predictions were frequently incorrect.

\subsection{Impact on Final Framework}

The lessons from these baseline experiments guided our transition to Whisper, which offers pretrained audio representations and richer semantic conditioning. Whisper’s improved caption quality enabled our confidence head, temperature scaling, and semantic ECE computation to operate in a setting where reliability estimation is both measurable and meaningful. Although not part of the final model, the ConvNeXt and BART experiments provided essential insight into model capacity, training behavior, and the importance of strong audio grounding for calibrated caption generation.

\end{document}